\documentclass[twocolumn,  
showpacs,  
preprintnumbers,  
aps,  
prd,  
a4paper,  
nofootinbib,  
tightenlines,  
floats, floatfix  
]{revtex4}

\usepackage{amsmath}
\usepackage{amsfonts}
\usepackage{graphicx}

\newcommand{\nn}{\nonumber \\}
\newcommand{\bea}{\begin{eqnarray}}
\newcommand{\ena}{\end{eqnarray}}

\newcommand{\R}{\cal{R}}
\renewcommand{\a}{\alpha}

\renewcommand{\d}{\delta}
\newcommand{\D}{\Delta}
\newcommand{\e}{\epsilon}
\renewcommand{\o}{\omega}

\begin{document}

\preprint{BI-TP 2010/01}

\title{Slow-roll inflation with a Gauss-Bonnet correction}

\bigskip

\author{Zong-Kuan Guo}
\email{guozk@physik.uni-bielefeld.de} \affiliation{Fakult{\"a}t
f{\"u}r Physik, Universit{\"a}t Bielefeld, Postfach 100131, 33501
Bielefeld, Germany}

\author{ Dominik J. Schwarz}
\email{dschwarz@physik.uni-bielefeld.de} \affiliation{Fakult{\"a}t
f{\"u}r Physik, Universit{\"a}t Bielefeld, Postfach 100131, 33501
Bielefeld, Germany}

\date{\today}

\begin{abstract}

We consider slow-roll inflation for a single scalar field with an
arbitrary potential and an arbitrary nonminimal coupling to the
Gauss-Bonnet term. By introducing a combined hierarchy of Hubble and
Gauss-Bonnet flow functions, we analytically derive the power
spectra of scalar and tensor perturbations. The standard consistency
relation between the tensor-to-scalar ratio and the spectral index
of tensor perturbations is broken. We apply this formalism to a
specific model with a monomial potential and an inverse monomial
Gauss-Bonnet coupling and constrain it by the 7-year Wilkinson
Microwave Anisotropy Probe data. The Gauss-Bonnet term with a
positive (or negative) coupling may lead to a reduction (or
enhancement) of the tensor-to-scalar ratio and hence may revive the
quartic potential ruled out by recent cosmological data.
\end{abstract}


\maketitle

\section{Introduction}
Inflation in the early Universe has become the standard model for
the generation of cosmological perturbations in the Universe, the
seeds for large-scale structure and temperature anisotropies of the
cosmic microwave background. The simplest scenario of
cosmological inflation is based upon a single, minimally coupled
scalar field with a flat potential. Quantum fluctuations of this
inflaton field give rise to an almost scale-invariant power spectrum
of isentropic perturbations (see Refs.~\cite{lyt99,bas05} for
reviews).

String theory is often regarded as the leading candidate for
unifying gravity with the other fundamental forces and for a quantum
theory of gravity. It is known that the effective supergravity action
from superstrings induces correction terms of
higher order in the curvature, which may play a
significant role in the early Universe. The simplest such correction
is the Gauss-Bonnet (GB) term in the low-energy effective action of
the heterotic string~\cite{gos87}. Such a term provides the
possibility of avoiding the initial singularity of the
Universe~\cite{ant93}. In the presence of an exponential potential
for the modulus field, nonsingular cosmological solutions were found
which begin in an asymptotically flat region, undergo
superexponential inflation and end with a graceful exit to a phase
with decreasing Hubble radius~\cite{tsu02}.

There are many works discussing accelerating cosmology with the GB
correction in four and higher
dimensions~\cite{guo07a,sat07,noj05,bro05}. Recently it has been
shown that the GB term might give rise to violent instabilities of
tensor perturbations~\cite{guo07}.  A model in which inflation is
driven by the GB term and a higher-order kinetic energy term was
studied. When the GB term dominates the dynamics of the background,
tensor perturbations exhibit violent negative instabilities around a
de Sitter background on small scales, in spite of the fact that
scale-invariant scalar perturbations can be achieved~\cite{guo07}.
Besides the kinetic and GB terms, a scalar potential arises
naturally from supersymmetry breaking or other nonperturbative
effects.

In a previous work, we investigated inflationary
solutions and resulting cosmological perturbations for the special
case of power-law inflation when both the GB
correction and the scalar potential are present~\cite{guo09}.
Power-law inflation happens when both the potential and
the GB coupling take an exponential form. In this model
instabilities of either scalar or tensor perturbations show up on
small scales for GB-dominated inflation. The GB correction
with a positive (or negative) coupling may lead to a reduction (or
enhancement) of the tensor-to-scalar ratio in the
potential-dominated case. This effect leads to tight constraints on
the magnitude of the GB correction from the Wilkinson Microwave
Anisotropy Probe (WMAP) 5-year analysis~\cite{kom09}.

Here we generalize our previous work to the more general case of
slow-roll inflation with an arbitrary potential and an arbitrary coupling.
Making use of a combined hierarchy ($\e_i$, $\d_i$) of Hubble and
GB flow functions (as defined below) with $|\e_i| \ll 1$ and
$|\d_i| \ll 1$, analogous  to the standard slow-roll approximation,
we derive the power
spectra of scalar and tensor perturbations.
In this scenario the spectral index of scalar perturbations contains
not only the Hubble flow parameters but also the GB flow parameters.
Moreover, the standard consistency relation of single-field slow-roll
inflation is modified.
In order to impose observational constraints on such models, we
focus on a specific model with a monomial potential and an inverse
monomial GB coupling.
We analyze the influence of the GB term on the scalar spectral index
$n_{\R}$ and the tensor-to-scalar ratio $r$.

This paper is organized as follows. In Sec. \ref{sect2} we define the
Hubble and GB flow functions. Then by using the background equations
of motion, we demonstrate that the slow-roll solution exists and is stable
under asymptotic conditions. In Sec. \ref{sect3} we calculate the power
spectra of scalar and tensor perturbations for the slow-roll
inflation. In Sec. \ref{sect4} our approach is applied to a specific
example. Section \ref{sect5} is devoted to conclusions.

\section{Slow-roll inflation with the GB correction \label{sect2}}
We consider the following action
\bea S = \int d^4x\sqrt{-g}
\left[\frac12 R - \frac{\o}{2}(\nabla \phi)^2
 - V(\phi)-\frac12\xi(\phi) R_{\rm GB}^2\right],
\label{action}
\ena
where $\phi$ is a scalar field with a potential
$V(\phi)$, $\o=\pm 1$, $R$ denotes the Ricci scalar, $R^{2}_{\rm GB}
= R_{\mu\nu\rho\sigma} R^{\mu\nu\rho\sigma} - 4 R_{\mu\nu}
R^{\mu\nu} + R^2$ is the GB term, and $\xi(\phi)$ is the GB
coupling. We work in Planckian units, $\hbar = c = 8\pi G = 1$. In a
spatially flat Friedmann-Robertson-Walker universe with scale factor
$a$, the background equations read
\bea
\label{beq1}
&& 6H^2 = \o \dot{\phi}^2 + 2V + 24\dot{\xi}H^3, \\
\label{beq12} && 2\dot{H} = -\o\dot{\phi}^2 + 4\ddot{\xi}H^2 +
4\dot{\xi} H
 \left(2\dot{H} - H^2\right), \\
&& \o \left(\ddot{\phi} + 3 H \dot{\phi}\right) + V_{,\phi} +12
\xi_{,\phi}
 H^2 \left(\dot{H}+H^2\right) = 0,
\label{beq2}
\ena
where a dot represents the time derivative, $(...)_{,\phi}$
denotes a derivative with respect to $\phi$,
and $H \equiv \dot{a}/a$ denotes the expansion rate. Since the GB
coupling is a function of $\phi$, one has
$\dot{\xi}=\xi_{,\phi}\dot{\phi}$ and
$\ddot{\xi}=\xi_{,\phi\phi}\dot{\phi}^2+\xi_{,\phi}\ddot{\phi}$.

Besides the slow-roll conditions $\dot{\phi}^2 \ll V$ and
$|\ddot{\phi}| \ll 3H|\dot{\phi}|$, well known for minimal-coupled single-field
inflation, we impose two extra conditions, namely
$4|\dot{\xi}|H \ll 1$ and
$|\ddot{\xi}| \ll |\dot{\xi}|H$. The background equations are
approximately given as
\bea
\label{seq1}
&& H^2 \simeq \frac13 V, \\
&& \dot{H} \simeq -\frac12 \o \dot{\phi}^2 - 2\dot{\xi}H^3, \\
&& \dot{\phi} \simeq -\frac{1}{3\o H}(V_{,\phi} + 12\xi_{,\phi}H^4),
\label{seq2}
\ena
which allows us to obtain the number of e-folds
\bea
\label{ne}
N(\phi) \simeq \int_{\phi_{\rm end}}^{\phi}
 \frac{3 \o V}{3 V_{,\phi}+4\xi_{,\phi}V^2}d\phi.
\ena

Following Ref.~\cite{sch01} we define a hierarchy of Hubble flow
parameters,
\bea
\e_1 = -\frac{\dot{H}}{H^2},
 \quad \e_{i+1} = \frac{d\ln|\e_i|}{d \ln a},
 \quad i \ge 1.
\ena
The expansion is accelerated as long as $\e_1<1$. In the
slow-roll approximation they can be related to the usual slow-roll
parameters. The new degrees of freedom introduced by the GB coupling
function $\xi(\phi)$ suggest to define an additional hierarchy of
flow parameters in the same way by
\bea
\d_1 = 4 \dot{\xi} H,
 \quad \d_{i+1} = \frac{d\ln|\d_i|}{d \ln a},
 \quad i \ge 1.
\ena
The slow-roll approximation becomes $|\e_i| \ll 1$ and $|\d_i| \ll 1$.

The definition of the Hubble and GB flow parameters renders
significant simplification in the involved expressions. From
Eqs.~(\ref{beq1}-\ref{beq2}) we can express the kinetic term and the
potential in terms of the flow parameters:
\bea
&& \o\dot{\phi}^2 = [2\e_1-\d_1(1+\e_1-\d_2)]H^2, \\
&& V = \frac12 [6-2\e_1+\d_1(-5+\e_1-\d_2)]H^2.
\ena
We see that the potential energy dominates over the kinetic energy
and the GB energy. During slow roll the sign of $\o$ is determined
by the sign of $(2\e_1-\d_1)$.
In the special case of $2\e_1=\d_1$, the field is frozen, which
corresponds to the constant Hubble parameter. We will not consider this
special case further.

It is known that slow roll is an attractor
that is rapidly approached by different initial conditions~\cite{lid00}.
Let us demonstrate that the slow-roll solution
(\ref{seq1}-\ref{seq2}) is the attractor of the system
(\ref{beq1}-\ref{beq2}) under the slow-roll condition.
From Eqs.~(\ref{beq12}) and (\ref{beq2}) one has
\bea
\label{hj1}
&& 2u\left(1+24\o\xi_{,\phi}\xi_{,\phi}H^4-4\xi_{,\phi}u H \right)
 H_{,\phi} = \nn
&& \hspace{6mm}
 - \o u^2 + 4\xi_{,\phi\phi}u^2H^2 \nn
&& \hspace{6mm}
 - 4\xi_{,\phi}H^2\left(4uH + \o V_{,\phi} + 12\o \xi_{,\phi}H^4\right),\\
&& u\left(1+24\o\xi_{,\phi}\xi_{,\phi}H^4-4\xi_{,\phi}u H \right)
 u_{,\phi} = \nn
&& \hspace{6mm}
 -3u H - \o \left(1-4\xi_{,\phi}u H\right) V_{,\phi}
 + 6\xi_{,\phi}H^2\big(3u^2 \nn
&& \hspace{6mm}
 - 4\o\xi_{,\phi\phi}u^2H^2
 -2\o H^2 + 12\o \xi_{,\phi} u H^3\big),
\ena
subject to the Friedmann constraint equation
\bea
6H^2 = \o u^2 + 2 V + 24 \xi_{,\phi}uH^3,
\label{hj2}
\ena
where $u = \dot{\phi}$.
Suppose $\bar{H}(\phi)$ and $\bar{u}(\phi)$ is the slow-roll
solution to the system (\ref{hj1}-\ref{hj2}). Add to this a linear
homogeneous perturbation $\d H(\phi)$ and $\d u(\phi)$; the
attractor condition will be satisfied if it becomes small as the
Universe expands. Inserting $H(\phi)=\bar{H}(\phi)+\d H(\phi)$ and
$u(\phi)=\bar{u}(\phi)+\d u({\phi})$ into Eqs.~(\ref{hj1}-\ref{hj2}),
we find that the linear perturbations satisfy
\bea
\d H_{,\phi} &=& -\frac{3H}{u}\left[1 +
 \frac{\d_1\e_1}{2\e_1-\d_1}
 + {\cal O}(\d_1\e_1,\d_1\d_2)\right]\d H, \\
\d u_{,\phi} &=& -\frac{3H}{u}\bigg[1 +
 \frac{2\e_1\e_2-8\e_1\d_1-\d_1\d_2-8\d_1^2}{6(2\e_1-\d_1)} \nn
&& + {\cal O}(\d_1\e_1,\d_1\d_2)\bigg]\d u,
\ena
which have an approximately decaying solution with $\d H
\propto \exp(-3N)$ and $\d u \propto \exp(-3N)$
if the Hubble and GB flow parameters vary slowly,
and hence all linear perturbations die away exponentially fast as
the number of e-folds increases.

\section{Power spectra \label{sect3}}
At linear order in perturbation theory, the Fourier modes of
curvature perturbations satisfy~\cite{car01}
\bea
v'' + \left(c_{\R}^2 k^2 - \frac{z''_{\R}}{z_{\R}}\right)v = 0,
\label{spm}
\ena
where a prime represents a derivative with respect
to conformal time $\tau = \int a^{-1}dt$, and where $z_{\R}$ and
$c_{\R}$ are given by
\bea
&& z_{\R}^2 = \frac{a^2(\o\dot{\phi}^2 +
 6 \D \dot{\xi}H^3)}{(1 - \frac12 \D)^2 H^2}, \\
&& c_{\R}^2 = 1 + \frac{8 \D \dot{\xi} H \dot{H} +
 2 \D^2 H^2 (\ddot{\xi}-\dot{\xi}H)}{\o\dot{\phi}^2+6\D\dot{\xi}H^3}
\label{cr2}
\ena
with $\D \equiv 4\dot{\xi}H/(1-4\dot{\xi}H)$. One
can express $z_{\R}^2$ and $c_{\R}^2$ in terms of the Hubble and GB
flow parameters,
\bea
&& z_{\R}^2 = a^2 \frac{F}{(1-\frac12 \D)^2}, \\
&& c_{\R}^2 = 1 - \D^2 \frac{2\e_1+\frac12 \d_1(1-5\e_1-\d_2)}{F},
\ena
where $\D = \d_1/(1-\d_1)$ and $F \equiv
2\e_1-\d_1(1+\e_1-\d_2)+\frac32 \D \d_1$. The effective mass term in
the scalar mode equation~(\ref{spm}) reads
\bea
\label{zrpp}
\frac{z_{\R}''}{z_{\R}} &=& a^2H^2 \Bigg[2 - \e_1
 +\frac32 \frac{\dot{F}}{H F}
 +\frac32 \frac{\dot{\D}}{H(1-\frac12 \D)} \nn
&&
 +\frac12 \frac{\ddot{F}}{H^2F}
 +\frac12 \frac{\ddot{\D}}{H^2(1-\frac12 \D)}
 -\frac14 \frac{\dot{F}^2}{H^2F^2} \nn
&&
 +\frac12 \frac{\dot{\D}^2}{H^2(1-\frac12 \D)^2}
 +\frac12 \frac{\dot{\D}}{H(1-\frac12 \D)}\frac{\dot{F}}{H F}
 \Bigg],
\ena
with
\bea
\frac{\dot{F}}{H} &=& \e_1\e_2(2-\d_1) - \d_1\d_2(1+\e_1-\d_2-\d_3) \nn
&& +\frac32 \D \d_2(\D+\d_1), \nn
\frac{\dot{\D}}{H} &=& \D^2 \frac{\d_2}{\d_1}, \nn
\frac{\ddot{F}}{H^2} &=& \e_1\e_2(-\e_1+\e_2+\e_3)(2-\d_1)
 + \e_1 \d_1\d_2(1+\e_1 \nn
&& -2\e_2-\d_2-\d_3) - \d_1\d_2^2(1+\e_1-\d_2-\d_3) \nn
&& - \d_1\d_2\d_3(1+\e_1-2\d_2-\d_3-\d_4) \nn
&& + \frac32 \D\d_2(\D+\d_1)(-\e_1+\Delta\frac{\d_2}{\d_1}+\d_3) \nn
&& + \frac32 \D\d_2(\D^2\frac{\d_2}{\d_1}+\d_1\d_2), \nn
\frac{\ddot{\D}}{H^2} &=& \D^2 \frac{\d_2}{\d_1}
 (-\e_1+2\D\frac{\d_2}{\d_1}-\d_2+\d_3). \nonumber
\ena

The Fourier modes of tensor perturbations satisfy~\cite{car01}
\bea
u'' + \left(c_{T}^2 k^2 - \frac{z''_{T}}{z_{T}}\right)u = 0,
\label{tpm}
\ena
where
\bea
&& z_{T}^2 = a^2(1 - 4\dot{\xi}H) , \\
&& c_{T}^2 = 1 - \frac{4(\ddot{\xi}-\dot{\xi}H)}{1 - 4\dot{\xi}H}.
\label{ct2}
\ena
Note that the coupling $\xi$ appears not only in
the $k^2$ term responsible for subhorizon oscillations but also in
the effective mass term $z''_{T}/z_{T}$. This differs from
k-inflation in which the equations of motion and evolution of the
tensor perturbations are not affected by nonminimal kinetic terms.
In terms of the Hubble and GB flow parameters $z_{T}^2$ and
$c_{T}^2$ can be written as
\bea
&& z_{T}^2 = a^2 (1-\d_1), \\
&& c_{T}^2 = 1 + \D (1-\e_1-\d_2).
\ena
The effective mass term in the tensor mode equation~(\ref{tpm}) reads
\bea
\label{ztpp}
\frac{z_{T}''}{z_{T}} &=& a^2H^2
 \bigg[2-\e_1-\frac32 \D\d_2-\frac12\D\d_2(-\e_1+\d_2+\d_3) \nn
&& -\frac14\D^2\d_2^2 \bigg].
\ena

If both $\e_1$ and $\d_1$ are constants, which corresponds to the
power-law inflation with an exponential potential and an exponential
GB coupling~\cite{guo09}, then Eqs.~(\ref{zrpp}) and (\ref{ztpp})
become
\bea
\frac{z_{\R}''}{z_{\R}} = \frac{z_{T}''}{z_{T}} =
 \frac{1}{\tau^2} \frac{2-\e_1}{(1-\e_1)^2}.
\ena
The spectral indices of scalar and tensor perturbations read
exactly
\bea
n_{\R}-1 = n_{T} = -\frac{2\e_1}{1-\e_1},
\ena
which are consistent with the results in Ref.~\cite{guo09}.
We note that only $\e_1$ appears in the spectral indices whether
for potential-dominated or GB-dominated inflation.

In general the Hubble and GB flow parameters are functions of
cosmic time. We shall assume that time derivatives of the
flow parameters can be neglected during slow-roll inflation, which
will allow us to obtain the leading contribution to the slow-roll
approximation. Under
this assumption one has $\tau^{-1} \simeq - aH(1-\e_1)$ and $\tau^2
z''_{\R}/z_{\R} \equiv \nu^2_{\R}-1/4$ can be approximated to
be constant. Then the general solution to Eq.~(\ref{spm}) is a linear
combination of Hankel functions
\bea
v = \frac{\sqrt{\pi |\tau|}}{2} e^{i(1+2\nu_{\R})\pi/4}
 \left[c_1H^{(1)}_{\nu_{\R}}(c_{\R}k|\tau|)
 + c_2 H^{(2)}_{\nu_{\R}}(c_{\R}k|\tau|)\right].
\ena
We choose $c_1=1$ and $c_2=0$, so that the usual Minkowski
vacuum state is recovered in the asymptotic past ($c_{\R}k|\tau| \to
\infty$). The power spectrum of curvature perturbations ${\cal
P}_{\R}=k^3|v/z_{\R}|^2/2\pi^2$ on the large scales ($c_{\R}k \ll
aH$) is
\bea
&& {\cal P}_{\R} = \frac{c_{\R}^{-3}}{|F|}
 \frac{H^2}{4\pi^2}
 \left(\frac{1-\D/2}{a H|\tau|}\right)^2
 \frac{\Gamma^2(\nu_{\R})}{\Gamma^2(3/2)}
 \left(\frac{c_{\R}k|\tau|}{2}\right)^{3-2\nu_{\R}} \nn
&& \hspace{7mm}
 \simeq \frac{2^{2\nu_{\R}-3}c_{\R}^{-3}}{|F|}
 \frac{H^2}{4\pi^2}
 \frac{\Gamma^2(\nu_{\R})}{\Gamma^2(3/2)}\bigg|_{c_{\R}k=aH},
\label{pss}
\ena
with spectral index
\bea
n_{\R} - 1 = 3-2\nu_{\R}.
\ena

As in the case of scalar perturbations, the power spectrum of tensor
perturbations ${\cal P}_{T} = 2k^3|2u/z_{T}|^2/2\pi^2$ is given by
\bea
&& {\cal P}_{T} =  \frac{8c_{T}^{-3}}{1-\d_1}
 \frac{H^2}{4\pi^2}
 \left(\frac{1}{a H|\tau|}\right)^2
 \frac{\Gamma^2(\nu_{T})}{\Gamma^2(3/2)}
 \left(\frac{c_Tk|\tau|}{2}\right)^{3-2\nu_{T}} \nn
&& \hspace{7mm}
 \simeq 2^{2\nu_{T}}c_{T}^{-3}
 \frac{H^2}{4\pi^2}
 \frac{\Gamma^2(\nu_{T})}{\Gamma^2(3/2)}\bigg|_{c_{T}k=aH},
\ena
with spectral index
\bea
n_{T} = 3-2\nu_{T},
\ena
where we have
defined $\nu^2_T \equiv \tau^2z''_T/z_T+1/4$. All background
quantities above are evaluated at the moment such that $c_Tk=aH$.
This is not exactly the same time as the horizon-crossing time in
Eq.~(\ref{pss}) for scalar modes, but to lowest order in the
slow-roll parameters this difference is unimportant. We can use the
slow-roll approximation to estimate the amount of e-folds between
horizon crossing of the scalar mode and the tensor mode with a
reference scale $k$, $\D N \sim \ln (c_{T}/c_{\R}) \sim \d_1/2$. An
important observational quantity is the tensor-to-scalar ratio which
is defined as
\bea
\label{ttsr}
r \equiv \frac{{\cal P}_{T}}{{\cal P}_{\R}}
 \simeq 2^{3+2\nu_{T}-2\nu_{\R}} |F|
 \frac{c_{\R}^3}{c_{T}^3}
 \frac{\Gamma^2(\nu_T)}{\Gamma^2(\nu_{\R})}.
\ena

To first order in the slow-roll approximation, we have
\bea
&& c_{\R}^2 \simeq 1 - \frac{\d_1^2(4\e_1 + \d_1)}{2(2\e_1-\d_1)}, \\
&& \frac{z_{\R}''}{z_{\R}} = a^2 H^2\bigg[2-\e_1
 +\frac{3(2\e_1\e_2-\d_1\d_2)}{2(2\e_1-\d_1)} \nn
&& \hspace{9mm}+{\cal O}(\e_1\e_2,\d_1\d_2)\bigg],\\
&& c_{T}^2 \simeq 1 + \d_1, \\
&& \frac{z_{T}''}{z_{T}} = a^2 H^2 \left[2-\e_1+{\cal O}(\d_1\d_2)\right].
\ena
The spectral indices of scalar and tensor perturbations read
\bea
\label{sis}
&& n_{\R}-1 \simeq -2\e_1 - \frac{2\e_1\e_2-\d_1\d_2}{2\e_1-\d_1},\\
&& n_{T} \simeq -2\e_1,
\label{sit}
\ena
which show that the spectral index of
scalar perturbation contains not only the Hubble flow parameters but
also the GB flow parameters. Even for a solution very close to de
Sitter inflation (i.e., $\e_i \approx 0$), the GB term can lead
to a red ($\d_2>0$) or blue
($\d_2<0$) power spectrum of scalar perturbation. If $|\d_1| \ll
\e_1$, the spectral indices are the same as for a potential-driven
slow-roll inflation.

The tensor-to-scalar ratio~(\ref{ttsr}) is approximately
\bea
\label{ratio}
r \simeq 8|2\e_1-\d_1| \neq -8n_T,
\ena
which is the modified consistency relation. The degeneracy of standard
consistency relation is broken in the slow-roll inflation with the
Gauss-Bonnet correction. For this reason, the future experimental
checking of this relation is usually regarded as an important test
of the simplest forms of inflation.

The Hubble and GB flow parameters can be expressed in terms of the
potential and the GB coupling
\bea
\e_1 &\simeq& \frac{Q}{2} \frac{V_{,\phi}}{V},\\
\e_2 &\simeq& -Q \left(\frac{V_{,\phi\phi}}{V_{,\phi}}
 - \frac{V_{,\phi}}{V} + \frac{Q_{,\phi}}{Q}\right), \\
\d_1 &\simeq& -\frac43 \xi_{,\phi} Q V ,\\
\d_2 &\simeq& -Q \left(\frac{\xi_{,\phi\phi}}{\xi_{,\phi}} +
\frac{V_{,\phi}}{V}
 + \frac{Q_{,\phi}}{Q}\right),
\ena
where $Q \equiv \o(V_{,\phi}/V + 4\xi_{,\phi}V/3)$.

The key result of our paper is the general slow-roll expression
for GB inflation, Eqs.~(\ref{sis}), (\ref{sit}) and (\ref{ratio}),
which is new and follows from a nontrivial calculation.

\section{An example model \label{sect4}}
Let us consider a specific inflation model
\bea
\label{poco}
V(\phi) = V_0 \phi^n, \quad \xi(\phi) = \xi_0 \phi^{-n}.
\ena
This potential has been widely studied. The specific choice of GB
coupling allows us to find an analytic relation between
the spectral index of curvature perturbations and the
tensor-to-scalar ratio. If $\a \equiv 4V_0\xi_0/3=1$, all flow
parameters vanish. The motion of the inflaton is frozen because the
force due to the slope of the potential is exactly balanced by one,
the slope of the GB coupling. In this case, exact de Sitter
inflation can be realized for the monomial potential and the inverse
monomial GB coupling. If $\a<1$, choosing $\o=1$ is required for a
positive $\e_1$.
In this case the contribution of the positive GB term increases
the Hubble expansion rate during inflation, which makes the evolution
of the inflaton slower than in the case of standard slow-roll inflation,
while the contribution of the negative GB term decreases the Hubble
expansion rate. If $\a>1$,
we choose $\o=-1$ to guarantee $\e_1>0$. The potential force drives
the inflaton to climb up the potential while the GB force drives the
field to roll down. Since the GB force dominates over the
potential force, slow-roll inflation can be realized. In what follows
we restrict our discussion to the case of $\a<1$.

\begin{figure}
\begin{center}
\includegraphics[width=8cm]{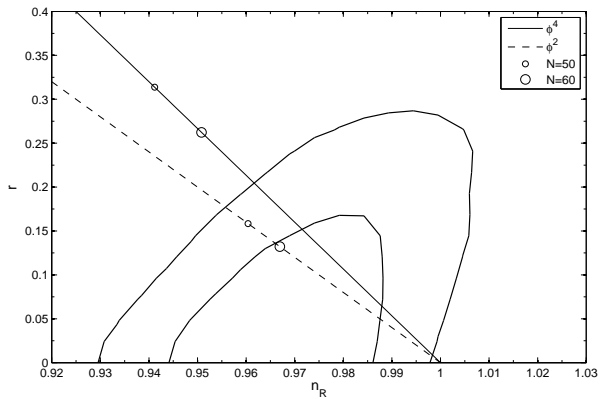}
\caption{Two-dimensional joint marginalized constraint (68\% and 95\% confidence level)
on the scalar spectral index $n_{\R}$ and the tensor-to-scalar ratio $r$
derived from the data combination of WMAP7+BAO+$H_0$ by imposing the
standard consistency relation.
The symbols show the predictions from the $\phi^4$-potential (solid line)
and $\phi^2$-potential (dashed line) models with the number of e-folds
equal to 50 (small) and 60 (large).}
\label{fig_wmap}
\end{center}
\end{figure}

The flow parameters are
\bea
\e_1 &\simeq& \frac12 n^2 (1-\a) \phi^{-2}, \\
\e_2 &\simeq& 2n (1-\a) \phi^{-2}, \\
\d_1 &\simeq& n^2 \a(1-\a) \phi^{-2}, \\
\d_2 &\simeq& 2n (1-\a) \phi^{-2}.
\ena
From Eqs.~(\ref{sis}) and (\ref{ratio}) one gets
\bea
n_{\R}-1 &=& -n(n+2)(1-\a)\phi^{-2}, \\
r &=& 8n^2(1-\a)^2\phi^{-2}.
\ena
Inflation ends at $\e_1(\phi_{\rm end})=1$, which gives the value
of the field at the end of inflation
\bea
\phi_{\rm end}^2 = \frac12 n^2(1-\a).
\ena
Then from~(\ref{ne})
we find the value of the field $N$ e-folds before the end of inflation
\bea
\phi^2 = 2n(1-\a)(N+\frac{n}{4}).
\ena
The spectral index $n_{\R}$
and the tensor-to-scalar ratio $r$ can be written in terms of the
function of $N$:
\bea
n_{\R}-1 &=& -\frac{2(n+2)}{4N+n}, \\
r &=& \frac{16n(1-\a)}{4N+n}.
\ena
Note that the spectral index is
independent of $V_0$ and $\xi_0$, but the tensor-to-scalar ratio
depends on $\a=4V_0\xi_0/3$. The GB correction leads to a reduction
of the tensor-to-scalar ratio if $\xi_0>0$ while an enhancement if
$\xi_0<0$, which is still valid in the power-law inflation
model with the exponential potential and GB coupling~\cite{guo09}.

\begin{figure}
\begin{center}
\includegraphics[width=8cm]{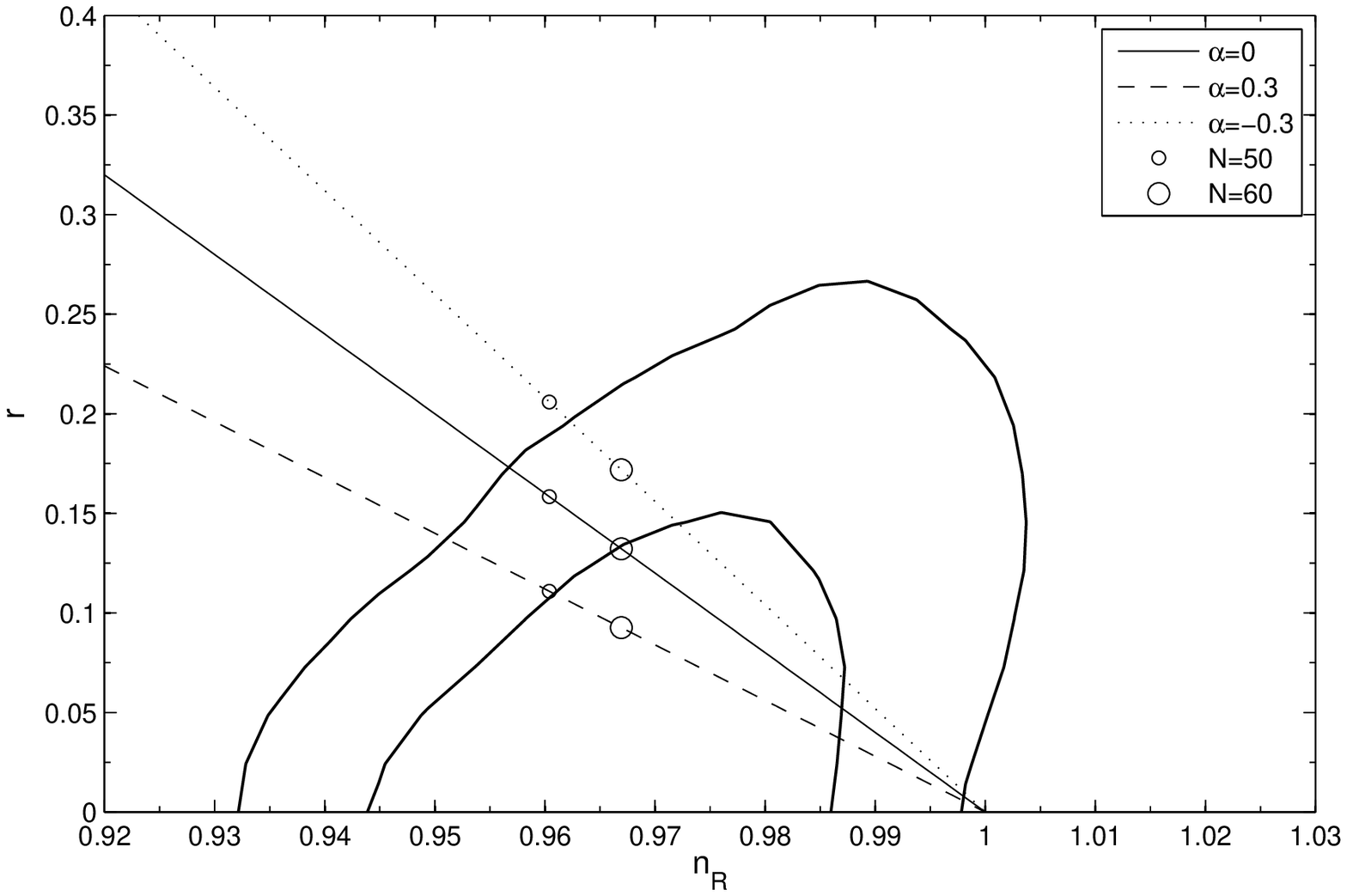}
\includegraphics[width=8cm]{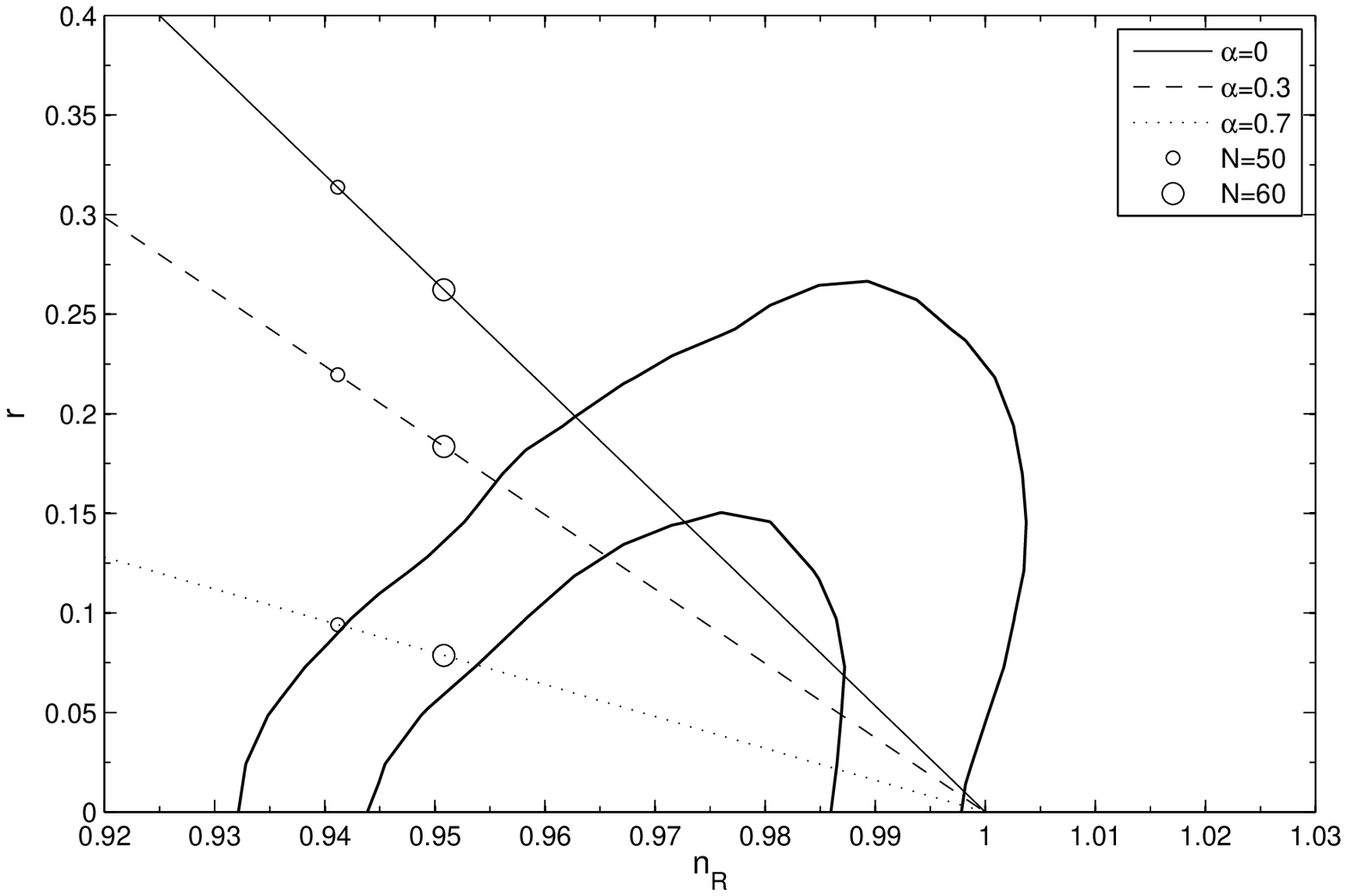}
\caption{Tensor-to-scalar ratio $r$ versus the spectral index
$n_{\R}$ for the inflation model (\ref{poco}) with $n=2$ (top panel)
and $n=4$ (bottom panel).
The contours show the 68\% and 95\% confidence level derived
from WMAP7+BAO+$H_0$ without the consistency relation.}
\label{fig-rn}
\end{center}
\end{figure}

Figure~\ref{fig_wmap} shows the two-dimensional joint
marginalized constraint (68\% and 95\% confidence level)
on $n_{\R}$ and $r$ from the 7-year WMAP+BAO+$H_0$ by imposing the
standard consistency relation~\cite{kom10}.
The symbols show the predictions from the $\phi^4$-potential (solid line)
and $\phi^2$-potential (dashed line) models with the number of e-folds
equal to 50 (small) and 60 (large).
We can see that the predicted points with $N=50, 60$ for the
quartic potential are far away from the 95\% region. The quadratic
potential is consistent with the data.

However, the consistency relation $n_T=-r/8$ is broken in the
slow-roll inflation with the GB correction.
Therefore, in our analysis, $n_T$ is varied independent of
the tensor-to-scalar ratio.
For the tensor perturbations we assume a power-law power spectrum,
with a uniform prior on $n_T$ as $-0.5<n_T<0$.
In Fig.~\ref{fig-rn} we show the $1\sigma$ and $2\sigma$ contours
derived from the data combination of WMAP7+BAO+$H_0$ by using
the CosmoMC package~\cite{lew02}.
Compared to the contours of Fig.~\ref{fig_wmap} we find that
the joint constraint on $n_{\R}$ and $r$ becomes a little tighter.
The WMAP7+BAO+$H_0$ data do not constrain $n_T$.
Basically all values allowed by the prior are also allowed by
the potential and the coupling.

In Fig.~\ref{fig-rn} we plot the values of $n_{\R}$ and $r$
in the models with $n=2$ (top panel) and $n=4$ (bottom panel)
for different values of $N$ and $\a$.
We can see that the model parameter $\a$ can
shift the predicted $r$ vertically for a fixed number of e-folds.
For $n=2$, the model with a positive $\a$ is more favored observationally.
For $n=4$, the model with $\a>0.7$ is consistent with the data within
the 95\% confidence level, in which the
prediction for the tensor-to-scalar ratio is smaller than
the $\a=0$ case while the prediction for $n_{\R}$ is the same
as the $\a=0$ case.
Other ways to avoid the exclusion of the $\phi^4$ potential have
been studied in Ref.~\cite{ram09}.

\section{Conclusions and discussions\label{sect5}}
In this paper we have studied slow-roll inflation with a nonminimally
coupled Gauss-Bonnet term. We have defined a combined
hierarchy ($\e_i, \d_i$) of Hubble and GB flow functions such that
$|\e_i| \ll 1$ and $|\d_i| \ll 1$ is the analogue of the standard
slow-roll approximation. It has been demonstrated that slow-roll
solution is the attractor solution under the slow-roll condition.
We have analytically derived the power spectra of scalar and tensor
perturbations. In general the spectral index of scalar perturbations
depends on the Hubble flow parameters and the GB flow
parameters. However, the spectral index of tensor perturbations
is independent of the GB flow parameters to first order in the slow-roll
approximation. In this scenario the standard consistency relation
does not hold because of the GB correction.

We apply our general formalism to large-field inflation
with a monomial potential and the GB coupling~(\ref{poco}).
We focus on the case of $\o=1$ and $\a<1$ since the field theory of
phantom-type fields encounters the problem of stability.
In this case, the GB term with the positive (or negative) coupling
slows down (or speeds up) the evolution of the inflaton during inflation,
which decreases (or increases) the energy scale of the potential
to be in agreement with the amplitude of scalar perturbations.
However the amplitude of tensor perturbations only depends on
the energy scale of the potential at the horizon-crossing time.
Therefore, the tensor-to-scalar ratio is suppressed for $\a>0$ while
it is enhanced for $\a<0$.

As shown in Fig.~\ref{fig-rn}, the model parameter $\a$ can shift
the predicted $r$ vertically for a fixed number of e-folds in the
$n_{\R}$-$r$ plane.
For $n=2$, the quadratic potential can be made a better fit to
the data by the positive GB coupling.
For $n=4$, it is known that the model with $\a=0$ is excluded by
the WMAP7+BAO+$H_0$ analysis.
However, in our scenario of inflation $\a>0.7$ is within the $2\sigma$
contour for $N>50$, and
it is consistent with the data within the 95\% confidence level.

The results of this work are generic as soon as nonminimal couplings
are considered.
While it is always possible by means of a conformal transformation
to work in the Einstein frame and
to avoid the presence of a $\phi^2R$ term in the Lagrangian, the coupling
of the scalar field to the GB term cannot be argued away by the same
conformal transformation.
While we studied perturbation spectra in the Einstein frame,
similar properties hold in the Jordan frame.

\begin{acknowledgments}
We thank H.-T.~Ding and E.~Komatsu for useful discussions.
Our numerical analysis was performed on the HPC cluster of
the RWTH Aachen University.
This work was supported by the Alexander von Humboldt Foundation.
\end{acknowledgments}


\begin{thebibliography}{99}
\bibitem{lyt99}
V.~F.~Mukhanov, H.~A.~Feldman and R.~H.~Brandenberger,
  Phys.\ Rept.\ {\bf 215}, 203 (1992);
D.~H.~Lyth and A.~Riotto,
  Phys.\ Rept.\ {\bf 314}, 1 (1999)
  [arXiv:hep-ph/9807278];
K.~A.~Malik and D.~Wands,
  Phys.\ Rept.\ {\bf 475}, 1 (2009)
  [arXiv:0809.4944].
\bibitem{bas05}
J.~E.~Lidsey, A.~R.~Liddle, E.~W.~Kolb, E.~J.~Copeland,
  T.~Barreiro and M.~Abney,
  Rev.\ Mod.\ Phys.\  {\bf 69}, 373 (1997)
  [arXiv:astro-ph/9508078];
B.~A.~Bassett, S.~Tsujikawa and D.~Wands,
  Rev.\ Mod.\ Phys.\ {\bf 78}, 537 (2006)
  [arXiv:astro-ph/0507632].
\bibitem{gos87}
D.~J.~Grossa and J.~H.~Sloana,
  Nucl.\ Phys.\ B {\bf 291}, 41 (1987);
M.~Gasperini, M.~Maggiore and G.~Veneziano,
  Nucl.\ Phys.\ B {\bf 494}, 315 (1997)
  [arXiv:hep-th/9611039].
\bibitem{ant93}
I.~Antoniadis, J.~Rizos and K.~Tamvakis,
  Nucl.\ Phys.\ B {\bf 415}, 497 (1994)
  [arXiv:hep-th/9305025];
S.~Kawai, M.~a.~Sakagami and J.~Soda,
  Phys.\ Lett.\  B {\bf 437}, 284 (1998)
  [arXiv:gr-qc/9802033];
S.~Kawai and J.~Soda,
  Phys.\ Lett.\  B {\bf 460}, 41 (1999)
  [arXiv:gr-qc/9903017];
S.~Tsujikawa,
  Phys.\ Lett.\ B {\bf 526}, 179 (2002)
  [arXiv:gr-qc/0110124];
A.~Toporensky and S.~Tsujikawa,
  Phys.\ Rev.\ D {\bf 65}, 123509 (2002)
  [arXiv:gr-qc/0202067].
\bibitem{tsu02}
S.~Tsujikawa, R.~Brandenberger and F.~Finelli,
  Phys.\ Rev.\ D {\bf 66}, 083513 (2002)
  [arXiv:hep-th/0207228].
\bibitem{guo07a}
K.~Bamba, Z.~K.~Guo and N.~Ohta,
  Prog.\ Theor.\ Phys.\ {\bf 118}, 879 (2007)
  [arXiv:0707.4334];
K.~Andrew, B.~Bolen and C.~A.~Middleton,
  Gen.\ Rel.\ Grav.\ {\bf 39}, 2061 (2007)
  [arXiv:0708.0373];
R.~Chingangbam, M.~Sami, P.~V.~Tretyakov and A.~V.~Toporensky,
  Phys.\ Lett.\ B {\bf 661}, 162 (2008)
  [arXiv:0711.2122];
I.~V.~Kirnos and A.~N.~Makarenko,
  arXiv:0903.0083.
\bibitem{sat07}
M.~Satoh, S.~Kanno and J.~Soda,
  Phys.\ Rev.\  D {\bf 77}, 023526 (2008)
  [arXiv:0706.3585];
M.~Satoh and J.~Soda,
  JCAP {\bf 0809}, 019 (2008)
  [arXiv:0806.4594].
\bibitem{noj05}
S.~Nojiri, S.~D.~Odintsov and M.~Sasaki,
  Phys.\ Rev.\ D {\bf 71}, 123509 (2005)
  [arXiv:hep-th/0504052];
G.~Cognola, E.~Elizalde, S.~Nojiri, S.~D.~Odintsov and S.~Zerbini,
  Phys.\ Rev.\ D {\bf 73}, 084007 (2006)
  [arXiv:hep-th/0601008];
T.~Koivisto and D.~F.~Mota,
  Phys.\ Lett.\ B {\bf 644}, 104 (2007)
  [astro-ph/0606078];
S.~Tsujikawa and M.~Sami,
  JCAP {\bf 0701}, 006 (2007)
  [arXiv:hep-th/0608178];
T.~Koivisto and D.~F.~Mota,
  Phys.\ Rev.\ D {\bf 75}, 023518 (2007)
  [hep-th/0609155];
B.~ M.~Leith and I.~P.~Neupane,
  JCAP {\bf 0705}, 019 (2007)
  [arXiv:hep-th/0702002];
B.~C.~Paul and S.~Ghose,
  arXiv:0809.4131;
M.~R.~Setare and E.~N.~Saridakis,
  Phys.\ Lett.\ B {\bf 670}, 1 (2008)
  [arXiv:0810.3296];
J.~Sadeghi, M.~R.~Setare and A.~Banijamali,
  Phys.\ Lett.\ B {\bf 679}, 302 (2009)
  [arXiv:0905.1468];
J.~Sadeghi, M.~R.~Setare and A.~Banijamali,
  Eur.\ Phys.\ J.\ C {\bf 64}, 433 (2009)
  [arXiv:0906.0713].
\bibitem{bro05}
R.~A.~Brown, R.~Maartens, E.~Papantonopoulos and V.~Zamarias,
  JCAP {\bf 0511}, 008 (2005)
  [arXiv:gr-qc/0508116];
J.~H.~He, B.~Wang and E.~Papantonopoulos,
  Phys.\ Lett.\ B {\bf 654}, 133 (2007)
  [arXiv:0707.1180];
E.~N.~Saridakis,
  Phys.\ Lett.\ B {\bf 661}, 335 (2008)
  [arXiv:0712.3806].
\bibitem{guo07}
Z.~K.~Guo, N.~Ohta and S.~Tsujikawa,
  Phys.\ Rev.\ D {\bf 75}, 023520 (2007)
  [arXiv:hep-th/0610336].
\bibitem{guo09}
Z.~K.~Guo and D.~J.~Schwarz,
  Phys.\ Rev.\ D {\bf 80}, 063523 (2009)
  [arXiv:0907.0427].
\bibitem{kom09}
E.~Komatsu, {\it et al.},
  Astrophys.\ J.\ Suppl.\ {\bf 180}, 330 (2009)
  [arXiv:1001.4538].
\bibitem{sch01}
D.~J.~Schwarz, C.~A.~Terrero-Escalante and A.~A.~Carcia,
  Phys.\ Lett.\ B {\bf 517}, 243 (2001)
  [arXiv:astro-ph/0106020];
S.~M.~Leach, A.~R.~Liddle, J.~Martin and D.~J.~Schwarz,
  Phys.\ Rev.\ D {\bf 66}, 023515 (2002)
  [arXiv:astro-ph/0202094];
D.~J.~Schwarz and C.~A.~Terrero-Escalante,
  JCAP {\bf 0408}, 003 (2004)
  [arXiv:hep-ph/0403129].
\bibitem{lid00}
A.~R.~Liddle and D.~H.~Lyth, ``Cosmological inflation and large-scale structure'',
  Cambridge University Press, Cambridge, England (2000).
\bibitem{car01}
J.~Hwang and H.~Noh,
  Phys.\ Rev.\ D {\bf 61}, 043511 (2000)
  [arXiv:astro-ph/9909480];
C.~Cartier, J.~Hwang and E.~J.~Copeland,
  Phys.\ Rev.\ D {\bf 64}, 103504 (2001)
  [arXiv:astro-ph/0106197];
J.~Hwang and H.~Noh,
  Phys.\ Rev.\ D {\bf 71}, 063536 (2005)
  [arXiv:gr-qc/0412126].
\bibitem{kom10}
E.~Komatsu, {\it et al.},
  arXiv:1001.4538.
\bibitem{lew02}
A.~Lewis and S.~Bridle,
  Phys.\ Rev.\ D {\bf 66}, 103511 (2002)
  [arXiv:astro-ph/0205436].
\bibitem{ram09}
S.~Tsujikawa and B.~Gumjudpai,
  Phys.\ Rev.\ D {\bf 69}, 123523 (2004)
  [arXiv:astro-ph/0402185];
E.~Ramirez and D.~J.~Schwarz,
  Phys.\ Rev.\ D {\bf 80}, 023525 (2009)
  [arXiv:0903.3543].
\end{thebibliography}
\end{document}